\documentstyle[12pt]{article}

\begin{document}

\hfill{UTTG-02-03}

\vspace{36pt}

\begin{center}
{\bf Damping of Tensor Modes in Cosmology}\\

\vspace{12pt}

Steven Weinberg\\
{\em Theory Group, Physics Department, University of Texas, \\Austin, 
TX, 78712}
\end{center}

\vspace{12pt}

\footnotesize
An analytic formula is given for the traceless transverse part of the 
anisotropic stress tensor due to free streaming neutrinos, and used to 
derive an integro-differential equation for the propagation of 
cosmological gravitational waves.  The solution shows that anisotropic
stress reduces the squared amplitude by 35.6 \% for 
wavelengths that enter the horizon during the radiation-dominated 
phase, independent of any cosmological parameters.  This decreases the
tensor temperature and polarization correlation functions for these 
wavelengths 
by the same amount.  
The effect is less for wavelengths that enter the horizon at later 
times.
At the longest wavelengths the decrease in the tensor correlation 
functions
due to neutrino free streaming ranges from 10.7\% for  
$\Omega_Mh^2=0.1$ to 9.0\% for $\Omega_Mh^2=0.15$.  An Appendix gives 
a general proof that tensor as well as 
scalar modes satisfy a conservation law for perturbations outside the 
horizon, even when the 
anisotropic stress tensor is not negligible.

\normalsize

\vspace{12pt}

\begin{center}
{\bf I. Introduction}
\end{center}

\vspace{12pt}

It is widely expected that the observation of cosmological tensor 
fluctuations through measurements of the polarization of the microwave 
background 
may provide a uniquely valuable check on the validity of simple
inflationary cosmologies.  For instance,  for a large class of 
inflationary theories with single scalar fields satisfying the
``slow roll'' approximation, the wave-number dependence 
${\cal P}_S\propto k^{n_S-1}$
and ${\cal P}_T\propto k^{n_T}$ of the scalar and tensor power 
spectral functions and
the ratio of these spectral functions after horizon exit during 
inflation are related by[1]
\begin{equation}
 {\cal P}_T/{\cal P}_S=-n_T/2 \;.
\end{equation} But in order to use observations to check such 
relations, we need to know what 
happens to the fluctuations between the time of inflation and the 
present.  There is
a very large literature on the scalar modes, but ever since the first 
calculations[2] of 
the production of tensor modes in inflation, with only one 
exception[3] known to me, the interaction of these 
modes with matter and radiation has simply been assumed to be 
negligible
in studies of the 
cosmic microwave background[4].  It is 
not included in the widely used computer program of Seljak and 
Zaldarriaga[5].  As we shall see, the effect is not negligible even at 
the relatively low values of $\ell$ where the $B$-type polarization 
multipole coefficients $C_{B\,\ell}$ are likely to be first
measured, and becomes quite significant for larger values of $\ell$.

\vspace{12pt}

\begin{center}
{\bf II. Damping Effects in the Wave Equation}
\end{center}

\vspace{12pt}

The interaction of tensor modes with matter and radiation vanishes in 
the case 
of perfect fluids, but not in the presence of traceless transverse 
terms in the anisotropic stress tensor.  In general, the tensor 
fluctuation satisfies 
\begin{equation}
\ddot{h}_{ij}+\left(\frac{3\dot{a}}{a}\right)\dot{h}_{ij}-\left
(\frac{\nabla^2}{a^2}\right)h_{ij}=16\pi G\pi_{ij}\;,
\end{equation}
where dots indicate ordinary time derivatives.
Here the components of the perturbed metric are 
\begin{equation}
g_{00}=-1\;,~~~~g_{i0}=0\;,~~~~g_{ij}({\bf 
x},t)=a^2(t)\,\Big[\delta_{ij}+ h_{ij}({\bf x},t)\Big]
\end{equation} 
where $h_{ij}({\bf x},t)$  is treated as a small perturbation; and
$\pi_{ij}({\bf x},t)$ is the anisotropic part of the stress tensor, 
defined by
writing the spatial part of the perturbed energy-momentum tensor as $
T_{ij}=\bar{p}\,g_{ij}+a^2\pi_{ij}\;,$
or equivalently
\begin{equation} 
T^i{}_j=\bar{p}\,\delta_{ij}+\pi_{ij}\;,
\end{equation} 
where $\bar{p}$ is the unperturbed pressure.  In these formulas we are 
considering only tensor 
perturbations, so that
\begin{equation} 
h_{ii} 
=0\;,~~~\partial_ih_{ij}=0\;,~~~~~~\pi_{ii}=0\;,~~~\partial_i\pi_{ij}=
0\;.
\end{equation}   

For a perfect fluid $\pi_{ij}=0$, but this is
not true in general.  For instance, in any imperfect fluid with 
shear viscosity $\eta$, we have[6] $\pi_{ij}=-\eta \dot{h}_{ij}$.  
Nevertheless, as we shall show in the Appendix, even where 
hydrodynamic approximations are inapplicable,  $h_{ij}$ becomes time-
independent as  the wavelength of a mode leaves the horizon, and 
remains time-independent until horizon re-entry.  All modes of 
cosmological interest are still far outside the horizon at the 
temperature $\approx 10^{10}{}^\circ$K where neutrinos are going out 
of equilibrium with electrons and photons, so  $h_{ij}$ {\em can be 
effected by anisotropic inertia only later, when neutrinos are freely 
streaming.}

We can calculate the contribution of freely streaming neutrinos to 
$\pi_{ij}$ exactly[7].
We define a density $n({\bf x},{\bf p},t)$ as
\begin{equation}
n({\bf x},{\bf p},t)\equiv\sum_r \left(\prod_{i=1}^3\delta(x^i-
x^i_r(t))\right)
\left(\prod_{i=1}^3\delta(p_i-p_{ri}(t))\right)\;,
\end{equation} 
with $r$ labeling individual neutrino and antineutrino trajectories.  
The 
relativistic 
equations of motion in phase space for any metric with $g_{00}=-1$ and 
$g_{i0}=0$ are
\begin{equation} 
\dot{x}^i_r=\frac{p_r^i}{p_r^0}\;,~~~~\dot{p}_{ri}=\frac{p_r^jp_r^k}{2
p_r^0}\left(\frac{\partial g_{jk}}{\partial x^i}\right)_{x=x_r}\;.
\end{equation} 
It follows then that $n$ satisfies a Boltzmann equation
\begin{equation} 
\frac{\partial n}{\partial t}+\frac{\partial n}{\partial 
x^i}\frac{p^i}{p^0}+
\frac{\partial n}{\partial p_i}\frac{p^jp^k}{2p^0}\frac{\partial 
g_{jk}}{\partial x^i}=0\;,
\end{equation} 
it being understood that $p^i$ and $p^0$ are expressed in terms of the 
independent variable $p_i$ by $p^i=g^{ij}p_j$ and 
$p^0=(g^{ij}p_ip_j)^{1/2}$.  At a time $t_1$ soon after neutrinos 
started free streaming, $n$ had the ideal gas form (assuming zero 
chemical potentials)
\begin{equation} 
n({\bf x},{\bf 
p},t_1)=\frac{N}{(2\pi)^3}\left[\exp\left(\sqrt{g^{ij}({\bf 
x},t_1)p_ip_j}/k_{\cal B}T_1\right)+1\right]^{-1}\equiv n_1({\bf x}, 
{\bf 
p})\;,
\end{equation} 
where $N$ is the number of types of neutrinos, counting antineutrinos 
separately, and $ k_{\cal B}$ is Boltzmann's constant.  We therefore
write 
\begin{equation} 
n({\bf x}, {\bf p}, t)= n_1({\bf x},{\bf p})+ \delta n({\bf x}, {\bf 
p}, t)
\end{equation} 
so that $\delta n$ vanishes for $t=t_1$.

In the absence of metric perturbations, Eq.~(8) and the initial 
condition (9) have the 
solution $n({\bf p})=\bar{n}(p)$,
where $\bar{n}(p)$ is the zeroth-order part of $n_1$:
\begin{equation} 
\bar{n}(p)=\frac{N}{(2\pi)^3}\Big[\exp\left(p/ k_{\cal 
B}T_1a_1\right)+1\Big]^{-
1}\;,
\end{equation}
and $p\equiv \sqrt{p_ip_i}$.
To first order in metric perturbations, Eq.~(8) gives
\begin{equation} 
\frac{\partial \delta n({\bf x}, {\bf p}, t)}{\partial 
t}+\frac{\hat{p}_i}{a(t)}\frac{\partial \delta n({\bf x}, {\bf p}, 
t)}{\partial x^i}=-
\frac{p}{2a(t)}\bar{n}'(p)\hat{p}_i\hat{p}_j\hat{p}_k\frac{\partial}
{\partial x^k}\left(h_{ij}({\bf x}, t)-h_{ij}({\bf x}, 
t_1)\right)\;,
\end{equation}
where hats denote unit vectors.  (In putting the Boltzmann equation in 
this form, we use that fact that $n_1$ depends on ${\bf x}$ and $p_i$ 
only through the combination $g^{ij}({\bf x},t_1)p_ip_j$, so that to 
first order $\partial_k n_1({\bf x}, {\bf p})=-
p\,\bar{n}'(p)\hat{p}_i\hat{p}_j\partial_kh_{ij}({\bf x}, t_1)$.)

We now suppose that the ${\bf x}$-dependence of $h_{ij}({\bf x},t)$ is 
contained in a 
factor $\exp(i{\bf k}\cdot{\bf x})$, where ${\bf k}$ is a co-moving 
wave number.\footnote{Conventionally the co-moving coordinate ${\bf 
x}$ and wave number ${\bf k}$ are 
normalized by defining $a(t)$ so that $a=1$ at present.  Here we will 
leave this normalization arbitrary.}   Eq.~(12) and the initial 
condition that $\delta n=0$ at $t=t_1$ then have the solution
\begin{equation}
\delta n( {\bf p}, u)=-\frac{i }{2} 
p\,\bar{n}'(p)\,\hat{p}\cdot\hat{k}\,\hat{p}_i\,\hat{p}_j\int_{0}^u 
du'\;e^{i\hat{p}\cdot\hat{k}(u'-u)}\Big(h_{ij}(u')-h_{ij}( 
0)\Big)\;
\end{equation}  
where we now drop the position argument, and write $\delta n$ and 
$h_{ij} $ as functions of a variable $u$ instead of $t$, with $u$ 
defined 
as the wave number times the conformal time
\begin{equation}
u\equiv k\int_{t_1}^t\frac{dt'}{a(t')}\;.
\end{equation} 

The space part of the neutrino energy-momentum tensor is given by
\begin{equation}
T_\nu^i{}_j=\frac{1}{\sqrt{{\rm Det}g}}\sum_r \frac{p_r^i 
p_{rj}}{p_r^0}\delta^3({\bf x}-{\bf x}_r)=
\frac{1}{\sqrt{{\rm Det}g}}\int \left(\prod_{k=1}^3 dp_k\right) 
\frac{n\,p^i p_j}{p^0}
\end{equation} 
This yields terms of first order in $h_{ij}(u)$ from 
$p^i=g^{ij}p_j$ and $p^0=\sqrt{g^{ij}p_ip_j}$,
a term of first order in $h_{ij}(0)$ from the term $n_1$ in $n$, 
and a term of first order
in $h_{ij}(u)-h_{ij}(0)$ from $\delta n$.  Collecting all these 
terms and using Eq.~(5) yields a surprisingly simple formula for 
$\pi_{ij}$:
\begin{equation} 
\pi_{ij}(u)=-4\bar{\rho}_\nu(u)\,\int_0^u K(u-U)\,h'_{ij}(U)\,dU
\end{equation} 
where primes now indicate derivatives with respect to $U$ or $u$; $K$ 
is the kernel
\begin{equation} 
K(s)\equiv \frac{1}{16}\int_{-1}^{+1}dx\,(1-x^2)^2e^{i\,x\,s}=-
\frac{\sin s}{s^3}
-\frac{3\cos s}{s^4}+\frac{3\sin s}{s^5}\;,
\end{equation} 
and $\bar{\rho}_\nu=a^{-4}\int d^3p\;p\,\bar{n}(p)$ is the unperturbed 
neutrino energy density.

To continue, we use Eq.~(16) in Eq.~(2) 
and express time-derivatives in terms of $u$-derivatives.  This 
gives an integro-differential equation for $h_{ij}(u)$[8]:
\begin{equation} 
h_{ij}''(u)+\frac{2a'(u)}{a(u)} h_{ij}'(u)+ h_{ij}(u)=-24 
f_\nu(u)\left(\frac{a'(u)}{a(u)}\right)^2
\int_0^u K(u-U)\, h_{ij}'(U)\,dU\;,
\end{equation}  
where $f_\nu \equiv \bar{\rho}_\nu/\bar{\rho}$.  

We took the initial time $t_1$ to be soon after neutrinos started 
free streaming, so interesting perturbations are outside the horizon 
then, and for 
some time after.  As we show in the Appendix, $h_{ij}$
rapidly became time independent after horizon exit, and remained so 
until horizon re-entry.  In terms of
$u$, we then have the initial condition 
\begin{equation} 
h'_{ij}(0)=0\;.
\end{equation} 
The solution of Eq.~(18) can therefore be put in the general form
\begin{equation} 
h_{ij}(u)=h_{ij}(0)\chi(u)
\end{equation} 
where $\chi(u)$ satisfies the same integro-differential equation as 
$h_{ij}(u)$
\begin{equation} 
\chi''(u)+\frac{2a'(u)}{a(u)} \chi'(u)+\chi(u)=-24 
f_\nu(u)\left(\frac{a'(u)}{a(u)}\right)^2
\int_0^u K(u-U)\, \chi'(U)\,dU;,
\end{equation}
and the initial conditions
\begin{equation} 
\chi(0)=1\;,~~~~~~~~~~\chi'(0)=0\;.
\end{equation}

\vspace{12pt}

\begin{center}
{\bf III. Short Wavelengths}
\end{center}

\vspace{12pt}

We will first consider  wavelengths short enough to have re-entered 
the 
horizon during the radiation-dominated era (though long after neutrino 
decoupling), 
and then turn to the general case in the 
next section.
We can take the initial time $t_1$ to be early 
enough so that it can be approximated as $t_1\simeq 0$, with the zero 
of time defined so that during the radiation-dominated era we have 
$a\propto \sqrt{t}$.  Then in Eq.~(21) we can set $a'/a=1/u$, while 
for 3 neutrino flavors $f_\nu $ takes the constant value 
$f_\nu(0)=0.40523$.  Then Eq.~(21) becomes
\begin{equation} 
\chi''(u)+\frac{2}{u} \chi'(u)+\chi(u)=- \frac{24 
f_\nu(0)}{u^2}
\int_0^u K(u-U)\, \chi'(U)\,dU\;,
\end{equation}
Because of the 
decrease of the factor $1/u^2$, the right-hand of Eq.~(23) becomes 
negligible 
for $u\gg 1$, so deep inside the horizon 
the solution of Eqs.~(22) and (23)  approaches a homogeneous solution
\begin{equation} 
\chi(u)\rightarrow A\sin(u+\delta)/u
\end{equation} 
as compared with the solution $\sin(u)/u$ for $f_\nu=0$. A 
numerical solution of Eqs.~(22) and (23) shows that $\chi(u)$ follows 
the 
$f_\nu=0$ solution  pretty accurately until $u\approx 1$, when the 
perturbation enters the horizon, and thereafter rapidly approaches the 
asymptotic form (24), with 
$A=0.8026$ and $\delta $ very small.  This asymptotic form provides 
the initial condition for the later period when the matter energy 
density becomes first comparable and then greater than that of 
radiation, so the effect of neutrino damping at these later times is 
still only to reduce the tensor amplitude by the same factor 
$A=0.8026$.  Hence, for wavelengths that 
enter the horizon after electron--positron 
annihilation and well before radiation-matter equality, all quadratic 
effects
of the tensor modes in the cosmic microwave background, such as the 
tensor contribution to the temperature multipole 
coefficients $C_\ell$ and the whole of the ``B-B'' polarization 
multipole coefficients $C_{\ell\;B}$,  are 
35.6\% less than they would be without the damping due to free-
streaming neutrinos.   (Photons also contribute to $\pi_{ij} $, but 
this effect is 
much smaller because at last scattering photons contribute much less 
than 
40\% of the total energy.)

\vspace{12pt}

\begin{center}
{\bf IV. General Wavelengths}
\end{center}

\vspace{12pt}

To deal with perturbations that may enter the horizon after the matter 
energy density has become important, let us switch the independent 
variable from $u$ to $y\equiv a(t)/a_{\rm EQ}$,
where $a_{\rm EQ}$ is $a(t)$ at the time $t_{\rm EQ}$ of radiation-
matter equality.  To see how they are related, note that
\begin{equation} 
\frac{dy}{du}=\frac{\dot{a}}{ a_{\rm EQ}k/a}=\frac{a^2}{ a_{\rm 
EQ}k}H_0\sqrt{ 
\Omega_M\left(\frac{a_0}{a}\right)^3+(\Omega_\gamma+\Omega_\nu)\left(\
frac{a_0}{a}\right)^4}
\end{equation} 
The redshift of matter-radiation equality is given by $1+z_{\rm 
EQ}=a_0/a_{\rm EQ}=\Omega_M /(\Omega_\gamma+\Omega_\nu)$, so Eq.~(25) 
can be simplified to read
\begin{equation}
\frac{du}{dy}=\frac{Q}{\sqrt{1+y}}
\end{equation} 
where
\begin{equation}
Q\equiv \frac{k}{ a_0H_0\sqrt{\Omega_M(1+z_{\rm EQ})} }\;.
\end{equation} 
Since $u\rightarrow 0$ for $y\rightarrow 0$, the solution of Eq.~(26) 
is
\begin{equation}
u=2Q\Big(\sqrt{1+y}-1\Big)\;.
\end{equation} 
The Hubble constant at matter-radiation equality has the value $
H_{\rm EQ}=H_0\sqrt{2\Omega_M(1+z_{\rm EQ})^3}$, so Eq.~(27) can
be written
\begin{equation}
Q=\sqrt{2}k/k_{\rm EQ}\;,
\end{equation} 
where $ k_{\rm EQ}\equiv a_{\rm EQ}H_{\rm EQ}$ is the wave number of 
perturbations that just enter the horizon at the
time of radiation-matter equality.  (Hence in particular the results 
of the previous section apply for 
$Q\gg 1$.)

The fraction of the total energy density in neutrinos is well known
\begin{equation} 
f_\nu(y)=\frac{\Omega_\nu(a_0/a)^4}{\Omega_M(a_0/a)^3+(\Omega_\gamma+\
Omega_\nu)(a_0/a)^4}=\frac{f_\nu(0)}{1+y}
\end{equation} 
where 
\begin{equation} 
f_\nu(0)=\frac{\Omega_\nu}{\Omega_\nu+\Omega_\gamma}=0.40523\;.
\end{equation} 
A little algebra then lets us put Eq.~(21) in the form
\begin{equation} 
(1+y)\frac{d^2\chi(y)}{dy^2}+\left(\frac{2(1+y)}{y}+\frac{1}{2}\right)
\frac{d\chi(y)}{dy}+Q^2\chi(y)=
-\frac{24\,f_\nu(0)}{y^2}\int_0^y K(y,y')\frac{d\chi(y')}{dy'}dy'\;,
\end{equation} 
where $K(y,y')$ is the same as the $K(s)$ given by Eq.~(17), but with 
$s$ now given by
\begin{equation}
s\equiv z-z'=2Q\Big(\sqrt{1+y}-\sqrt{1+y'}\Big)
\end{equation} 
The initial conditions (22) now read
\begin{equation} 
\chi(0)=\left.\frac{d\chi(y)}{dy}\right|_{y=0}=0\;.
\end{equation}

We now have to face the complication that for general $Q$ the value of 
$y$ at last 
scattering is not in an asymptotic region where the effect of 
anisotropic inertia is 
simply to damp $\chi(t)$ by some constant factor.  We therefore now 
have to consider 
what feature of $\chi(t)$ is related to observations of the cosmic 
microwave background.
It is $\dot{\chi}$ that enters into the Boltzmann equation for 
perturbations to the temperature
and Stokes parameters[9], so in the approximation of a sudden 
transition from opacity to 
transparency, we expect the B-B and other multipole coefficients to 
depend on $\chi(y)$ only
through a factor
$|\chi'(y_L)|^2$, where $y_L=(1+z_{\rm EQ})/(1+z_L)$ is the value of 
$y$ at last scattering.
Hence we will be primarily
interested in calculating the value of $|\chi'(y_L)|^2$ for various 
values of $Q$,
and comparing these values with what they would be in the absence of 
anisotropic inertia.  

For $T_{\gamma 0}=2.725^\circ$K, we have 
$\Omega_\gamma+\Omega_\nu=4.15\times 10^{-5}h^{-2}$, so, taking
$1+z_L=1090$, the parameter $y_L$ is
$$ y_L=22.1\,\Omega_Mh^2 \;.$$  It will be useful also to have an idea 
of the value of $\ell$ for which
the multipole coefficients in various correlation functions are 
dominated by perturbations with
a given $Q$. The dominant contribution to a multipole coefficient of 
order $\ell$ comes from wave numbers
$k\simeq a_L\ell/d_L$, where $a_L$ is $a(t)$ at the time of last 
scattering, and $d_L$ is the angular diameter distance of the surface 
of last scattering, which for flat geometries is:
$$d_L=\frac{1}{H_0(1+z_L)}\int^1_{1/(1+z_L)}\frac{dx}{\sqrt{\Omega_Mx+
(1-\Omega_M)x^4}}\;,$$
where $z_L$ is the redshift of last scattering.  Thus the multipole 
order 
that receives its main contribution from wave lengths that are just 
coming into the horizon at
matter-radiation equality is
\begin{equation}
\ell_{\rm EQ}\equiv \frac{d_Lk_{\rm EQ}}{a_L}=\sqrt{2\Omega_M(1+z_{\rm 
EQ})}\int^1_{1/(1+z_L)}\frac{dx}{\sqrt{\Omega_Mx+(1-\Omega_M)x^4}}\;,
\end{equation} 
where $z_{\rm EQ}$ is the redshift of matter-radiation equality.  For 
present radiation temperature
$T_{\gamma 0}=2.725^\circ$K and $\Omega_Mh^2=0.15$ this redshift is 
$z_{\rm EQ}=3613$. If also $\Omega_M=0.3$ and $1+z_L=1090$  then the 
integral in Eq.~(35) has the value 3.195, and so Eq.~(35) gives 
$\ell_{\rm EQ}=149$.  Hence for these cosmological parameters, 
Eq.~(29) gives
$$Q=\sqrt{2}\frac{\ell}{\ell_{\rm EQ}}\simeq \frac{\ell}{105}\;.$$
When referring below to specific values of $\ell$, it will always be 
understood that the conversion from $Q$ to $\ell$ has been made using 
these cosmological parameters, but it should be kept in mind that the 
dependence of the
function $\chi(y)$ on $y$ and $Q$ is  independent of cosmological 
parameters, and that the value of $y$ at last scattering
depends only on $T_{\gamma 0}$, $1+z_L$. and $\Omega_Mh^2$, not on 
$\Omega_M$ or $\Omega_{\rm vac}$.

Let us first consider the case $Q\ll 1$, which for the above 
cosmological parameters corresponds to
$\ell\ll 100$.  Here the kernel $K(y,y')$ has the constant value 
$1/15$, and Eqs.~(32) and (34) have a 
solution of the form
\begin{equation}
\chi(y)\rightarrow 1-Q^2 g(y)~~~~~~{\rm for}~~Q\rightarrow 0
\end{equation} 
where $g(y)$ is independent of $Q$, and satisfies  the inhomogeneous 
differential equation
\begin{equation}
(1+y)\frac{d^2 
g(y)}{dy^2}+\left(\frac{2(1+y)}{y}+\frac{1}{2}\right)\frac{dg(y)}{dy}+
\frac{8f_\nu(0)}{5y^2}g(y)=1
\end{equation} 
and the initial conditions
\begin{equation}
g(0)=g'(0)=0\;.
\end{equation} 
According to the above discussion, the streaming of free neutrinos 
damps the various tensor correlation functions of the cosmic microwave 
background by a factor $|\chi'(y_L)/\chi'_0(y_L)|^2$, which for $Q\ll 
1$ becomes $|g'(y_L)/g'_0(y_L)|^2$, the subscript $0$ denoting 
quantities calculated ignoring this damping, i.e., for $f_\nu=0$, and 
$y_L$ again equal to the ratio of $a(t)$ at last scattering to that at 
matter-radiation equality.
Numerical solutions of Eqs.~(37) and (38) for $f_\nu(0)=0.40523$ and 
for $f_\nu=0$ show that the 
damping factor $|g'(y_L)/g'_0(y_L)|^2$ is very close to a linear 
function of $y_L$ and hence of $\Omega_Mh^2$ for observationally 
favored values of $\Omega_Mh^2$, increasing from 0.893 at 
$\Omega_Mh^2=0.10$ to 
0.910 for $\Omega_Mh^2=0.15$. 

This damping is relatively insensitive to $Q$ for small $Q$.  For 
instance, numerical integration
of Eqs.~(32) and (34) shows that for $\Omega_Mh^2=0.15$, the damping 
has only decreased from
9\% to 8\% for $Q=0.55$ ($\ell\simeq 58$), and to 7\% for $Q=0.8$ 
($\ell\simeq 84$).
Matters are more complicated for larger values of $Q$ and $\ell$, 
because the damping factor $|\chi'(y_L)/\chi'_0(y_L)|^2$
is the ratio of two oscillating functions with slightly different 
phases, so that the plot of $|\chi'(y_L)/\chi'_0(y_L)|^2$ {\it vs.}
$Q$ shows narrow spikes: this ratio becomes 
infinite at values of $Q$ for which $\chi'_0(y_L)$ vanishes and then 
almost immediately drops to zero at the slightly larger value of $Q$ 
for which $\chi'(y_L)$ vanishes.  (Even if we average over the small 
range 
of $y$ values over which last scattering occurs, the plot of 
$\langle|\chi'(y_L)/\chi'_0(y_L)|^2\rangle$ {\it vs.}
$Q$ still shows finite though high narrow spikes at the zeroes of 
$\chi'_0(y_L)$.)  These spikes are not particularly interesting, 
because they occur at values of $Q$  where $\chi'(y_L)$ is 
particularly small, so that the multipole coefficients in the various 
tensor
 temperature and polarization correlation functions will be very 
difficult to measure for the corresponding values of $\ell$.  The 
values of  $|\chi'(y_L)/\chi'_0(y_L)|^2$ in the relatively flat 
regions between the spikes steadily  decreases from the value $\simeq 
0.9$ for  $Q\ll 1$ to a value close to the result $.644$ found in the 
previous section for  $Q\simeq 10$.   

The effects considered in this paper will doubtless eventually be 
taken into account in the computer programs used to analyze data from 
PLANCK and other future facilities.  In the meanwhile, the planning of 
future observations should take into account that the damping of 
cosmological gravitational waves is not negligible.

\vspace{12pt}

\begin{center}
{\bf ACKNOWLEDGMENTS}
\end{center}

\vspace{12pt}

I am grateful for valuable conversations with Richard Bond, Lev 
Kofman, Eiichiro Komatsu, Richard Matzner and Matias Zaldarriaga.  
Thanks are  due to Michael Trott for advice regarding the numerical 
solution of Eq.~(18), and to Matthew Anderson for checking the 
numerical results.  
This research was supported in part by the Robert A. Welch Foundation,
by NSF Grant PHY-0071512, and by the US Navy Grant No. N00014-03-1-
0639, ``Quantum
Optics Initiative.''
\renewcommand{\theequation}{A\arabic{equation}}
\setcounter{equation}{0}

\vspace{12pt}

\begin{center}
{\bf APPENDIX: SUPERHORIZON CONSERVATION LAWS}
\end{center}

\vspace{12pt}

This Appendix will prove a result quoted in Section II, that in all 
cases there is a tensor mode whose amplitude remains constant outside 
the horizon, even where some particles may have mean free times 
comparable to the Hubble time.  The argument is similar to one used 
previously to show the existence under very general conditions of two 
scalar modes for which a quantity ${\cal R}$ is constant outside the 
horizon.[10]  It is based on the observation that for zero wave number 
the Newtonian gauge field equations and the dynamical equations for 
matter and radiation as well as the condition $k=0$ are invariant 
under  coordinate transformations that are {\em not} symmetries of the 
unperturbed metric.\footnote{In this respect, the theorem proved here 
is similar to the Goldstone theorem[11] of quantum field theory.  The 
modes for which ${\cal R}$ or $h_{ij}$ are constant outside the 
horizon take the place here of the Goldstone bosons that become free 
particles for long wavelength.}
The most general such transformations are
\begin{equation}
x^0\rightarrow x^0+\epsilon(t);,~~~~~x^i\rightarrow \left(\delta_{ij}-
\frac{1}{2}\omega_{ij}\right)x^j\;,
\end{equation} 
where $H\equiv \dot{a}/a$, $\epsilon(t)$ is an arbitrary function of 
time, and $\omega_{ij}=\omega_{ji}$ is an arbitrary constant matrix.  
Under these conditions we have something like a Goldstone theorem: 
since the metric satisfies the field equations both before and after 
the transformation,  the change in the metric under these 
transformations must also satisfy the field equations.  This change is 
simply 
\begin{equation} 
\delta g_{00}=\dot{\epsilon}(t)\;,~~~~~\delta g_{i0}=0\;,~~~~~~~\delta 
g_{ij}=a^2(t)\Big[-H(t)\epsilon(t)\delta_{ij}+\omega_{ij}\Big]\;.
\end{equation} 
This means that for zero wave number we always have a solution with 
scalar modes 
\begin{equation}
\Psi=H\epsilon-\omega_{ii}/3\;,~~~~~~\Phi=-\dot{\epsilon}
\end{equation} 
and a tensor mode
\begin{equation} 
h_{ij}=\omega_{ij}-\frac{1}{3}\delta_{ij}\omega_{kk}\;.
\end{equation} 
(The notation for $\Phi$ and $\Psi$ is standard, and the same as in 
Ref. [10].)  These are just gauge modes for zero wave number, but if 
they can be extended to non-zero wave number they become physical 
modes, since the transformations (A1) are not symmetries of the field 
equations except for zero wave number.  For the scalar modes there are 
field equations that disappear in the limit of zero wave number, so 
that conditions $\Phi=\Psi-8\pi G\pi_S$ and $\delta u=\epsilon$ (where 
$\pi_S$ is the scalar part of the anisotropic inertia, called $\sigma$ 
in Ref. [10])  and $\delta u$ is the perturbation to the velocity 
potential) must be imposed on the solutions (A3) for them to have an 
extension to non-zero wave number.  It follows then that the zero wave 
number scalar modes that become physical for non-zero wave number 
satisfy
\begin{equation} 
\dot{\epsilon}=-H\epsilon+\omega_{kk}/3-8\pi G\pi_S\;,~~~~~~\delta 
u=\epsilon\;.
\end{equation} 
Then  for zero wave number the quantity ${\cal R}\equiv -\Psi+h\delta 
u$ has the time-independent value 
\begin{equation} 
{\cal R}=\omega_{kk}/3;.
\end{equation} 
For tensor modes there are no field equations that disappear for zero 
wave number, so the solution $h_{ij}=$constant automatically has an 
extension to a physical mode for non-zero wave number.

As examples, we note that both the anisotropic stress tensor 
$\pi_{ij}=-\eta\dot{h}_{ij}$
for an imperfect fluid with shear viscosity $\eta$ and the tensor (16) 
for freely streaming neutrinos vanish for $\dot{h}_{ij}=0$, so in the 
limit of zero wave numbers Eq.~(2) has a solution 
with $\dot{h}_{ij}=0$.  The above theorem shows that this result 
applies even when some particle's mean free time is comparable with 
the Hubble time, in which case neither the hydrodynamic nor the free-
streaming approximations are applicable.

The solution with $\dot{h}_{ij}=0$ for zero wave number is not the 
only solution, but the other solutions decay rapidly after horizon 
exit.  There is no anisotropic inertia in scalar field theories, and 
in the absence of anisotropic inertia, Eq.~(2) for zero wave number 
has two solutions, one with $h_{ij}$ constant, and the other with 
$\dot{h}_{ij}\propto a^{-3}$, for which $h_{ij}$ rapidly becomes 
constant.  The energy-momentum tensor of the universe departs from the 
perfect fluid form later,
during neutrino decoupling, and perhaps also during reheating or 
periods of baryon or lepton nonconservation, but during all these 
epochs cosmologically interesting tensor fluctuations are far outside 
the horizon, and hence remain constant.

\begin{center}
References
\end{center}
\begin{enumerate}
\item A. Starobinsky, Sov. Astron. Lett. {\bf 11}, 133 (1985); E. D. 
Stewart
and D. H. Lyth, Phys. Lett. {\bf 302B}, 171 (1993).
\item V. A. Rubakov, M. Sazhin, and A. Veryaskin, 
Phys. Lett. {\bf 115B}, 189 (1982); R. Fabbri and 
M.D. Pollock, Phys. Lett. {\bf 125B}, 445 (1983); 
L. F. Abbott and M. B. Wise, Nuclear Physics {\bf B244}, 541 (1984).
A. A. Starobinskii, Sov. Astron. Lett. {\bf 11}, 133 (1985).
\item The effects of anisotropic inertia due to both neutrinos and 
photons were included in a large program of numerical calculations 
reported by J. R. Bond, in {\em Cosmology
and Large Scale Structure}, Les Houches Session LX, eds. R. Schaeffer, 
J. Silk, and
J. Zinn-Justin (Elsevier Science Press, Amsterdam, 1996).  Bond 
concluded from the numerical results that there is an average `$\sim$ 
20\%' reduction of the squared tensor amplitude for multipole order 
$\ell$ larger than about 100, and that this would not be observable in 
measurements of $C_\ell$ because according to Eq.~(1) tensor modes 
already
make a much smaller contribution to $C_\ell$ than scalar modes.  It is 
the 
prospect of cosmic microwave 
background polarization measurements that makes the effect of 
anisotropic inertia on the tensor amplitude important.
\item See, e.g., V. F. Mukhanov, H.A. Feldman, and R. H. 
Brandenberger, Physics 
Reports {\bf 215}, 203 (1992); M. S. Turner, M. White, and J. E. 
Lidsey, 
Phys. Rev. D {\bf 48}, 4613 (1993); M. S. Turner, Phys Rev. D. 
{\bf 55},  435 (1997); D. J. Schwarz, astro-ph/0303574.
\item U. Seljak and M. Zaldarriaga, Astrophys. J. {\bf 469}, 437 
(1996).
\item S. Weinberg, {\em Gravitation and Cosmology} (Wiley, New York, 
1972), Eq.~(15.10.39).  (It should be noted that $h_{ij}$ as defined 
in this reference is $a^2$ times the $h_{ij}$ used in the present 
work.)  For $k/a\gg \dot{a}/a$, this formula for $\pi $ gives the 
damping of gravitational waves that had been calculated by S. W. 
Hawking, Astrophys. J. {\bf 145}, 544 (1966).
\item Differential equations for both the 
scalar and tensor parts of the anisotropic stress tensor 
were given by J. R. Bond and A. S. 
Szalay, Astrophys. J. {\bf 274}, 443 (1983), using an orthonormal 
basis
instead of the coordinate basis used here, but the result was applied 
 only for the scalar modes. 
\item For perturbations outside the horizon, where $z\ll 1$, we can 
replace
$K(z-y)$ with $K(0)=1/15$, and the integral in Eq.~(8) becomes just 
$(h_{ij}(z)-h_{ij}(0))/15$.  Aside
from the term $h_{ij}(0)$, this equation in the radiation-dominated 
case
is then equivalent to Eq.~(4.3) of C. W. Misner, Astrophys. J. 
{\bf 151}, 431 (1968), which was derived to study a phenomenon 
different from that considered here: the approach to isotropy of a 
homogeneous anisotropic cosmology.  (This equation was  
generalized to the case of finite mean free times by C. Misner and R. 
Matzner, Astrophys. J. {\bf 171}, 415 (1972).)   Misner took 
$h_{ij}(0)=0$
(but $h_{ij}'(0)\neq 0$), on the ground that a constant term in 
$h_{ij}$ could be made to vanish by a coordinate transformation, 
and found a 
decaying solution.  But a constant term in $h_{ij}$ is only a gauge 
mode when $k$ is 
strictly zero.  As remarked in the Appendix, the existence of this 
gauge mode 
means that there is
a {\it physical} mode with $k\neq 0$ for which $h_{ij}$ becomes 
constant 
outside the horizon,
where $k$ is negligible, but which becomes time-dependent when the 
wavelength re-enters the 
horizon.  [After the preprint of this work was circulated, I learned 
of an
article by A. K. Rebhan and D. J. Schwarz, Phys. Rev. {\bf D 50}, 2541 
(1994),
which obtained an integro-differential equation like Eq.~(18), but 
with extra
terms representing more general initial conditions.  No attempt was 
made to 
identify the initial conditions that would actually apply 
cosmologically, or to
calculate the damping effect relevant to the cosmic microwave 
background.]
\item See, e. g., M. Zaldarriaga and U. Seljak, Phys. Rev. {\bf D55}, 
1830 (1997).
\item S. Weinberg, Phys. Rev. {\bf D67}, 123504 (2003).
\item J. Goldstone, Nuovo Cimento {\bf 9}, 154 (1961); J. Goldstone, 
A. Salam, and S.
Weinberg, Phys. Rev. {\bf 127}, 965 (1962).
\end{enumerate}
\end{document}